\documentclass[aps,prl,twocolumn,longbibliography,superscriptaddress]{revtex4-2}
\usepackage{bbm}
\usepackage{graphicx}
\usepackage{dcolumn}
\usepackage{bm}
\usepackage{subfigure}
\usepackage{amsmath}
\usepackage{feynmf}
\usepackage{hyperref}
\usepackage{CJK}
\usepackage{amssymb}
\usepackage{attachfile}

\newcommand{\bk}{\boldsymbol k}

\newcommand{\bq}{\boldsymbol q}

\newcommand{\bA}{\boldsymbol{A}}

\usepackage{braket}
\usepackage{esint}
\usepackage{times}

\begin{document}

\title{Light-Induced Even-Wave Spin Splittings in Nonmagnetic Centrosymmetric Systems \\ with Spin-Orbit Coupling }

\author{Xiao-Jiao Wang}
\affiliation{Guangdong Provincial Key Laboratory of Magnetoelectric Physics and Devices,
School of Physics, Sun Yat-sen University, Guangzhou 510275, China}

\author{Dongling Liu}
\affiliation{Guangdong Provincial Key Laboratory of Magnetoelectric Physics and Devices,
School of Physics, Sun Yat-sen University, Guangzhou 510275, China}

\author{Di Zhu}
\affiliation{Guangdong Provincial Key Laboratory of Magnetoelectric Physics and Devices,
	School of Physics, Sun Yat-sen University, Guangzhou 510275, China}

\author{Zheng-Yang Zhuang}
\affiliation{Guangdong Provincial Key Laboratory of Magnetoelectric Physics and Devices,
	School of Physics, Sun Yat-sen University, Guangzhou 510275, China}

\author{Zhongbo Yan}
\email{yanzhb5@mail.sysu.edu.cn}
\affiliation{Guangdong Provincial Key Laboratory of Magnetoelectric Physics and Devices,
School of Physics, Sun Yat-sen University, Guangzhou 510275, China}

\date{\today}

\begin{abstract}
Spin splitting underpins a vast range of spin-dependent phenomena. Traditionally, two primary mechanisms generate 
such splitting: relativistic spin-orbit coupling (SOC) and nonrelativistic magnetic exchange coupling (MEC). Governed by distinct symmetry constraints, they produce splittings of opposite parity---odd for SOC and even for MEC---a dichotomy that underpins the distinct spin physics of nonmagnetic and magnetic systems. In this work, we break this dichotomy by demonstrating the dynamic generation of even-parity spin splitting in centrosymmetric, nonmagnetic systems driven by circularly polarized light. We show that the symmetry of the induced splitting is controlled by the angular character of the underlying orbitals, enabling the realization of $s$-wave, $d$-wave, and $g$-wave spin-split band structures identical to those of ferromagnets and altermagnets. Furthermore, we find that these spin-split bands can naturally host a Chern insulator phase. We also discuss the associated spin and orbital magnetization. Our results establish a direct and previously unrecognized conceptual link between the two fundamental mechanisms of spin splitting. 
\end{abstract}

\maketitle

\begin{figure}[t]
\centering
\includegraphics[width=0.45\textwidth]{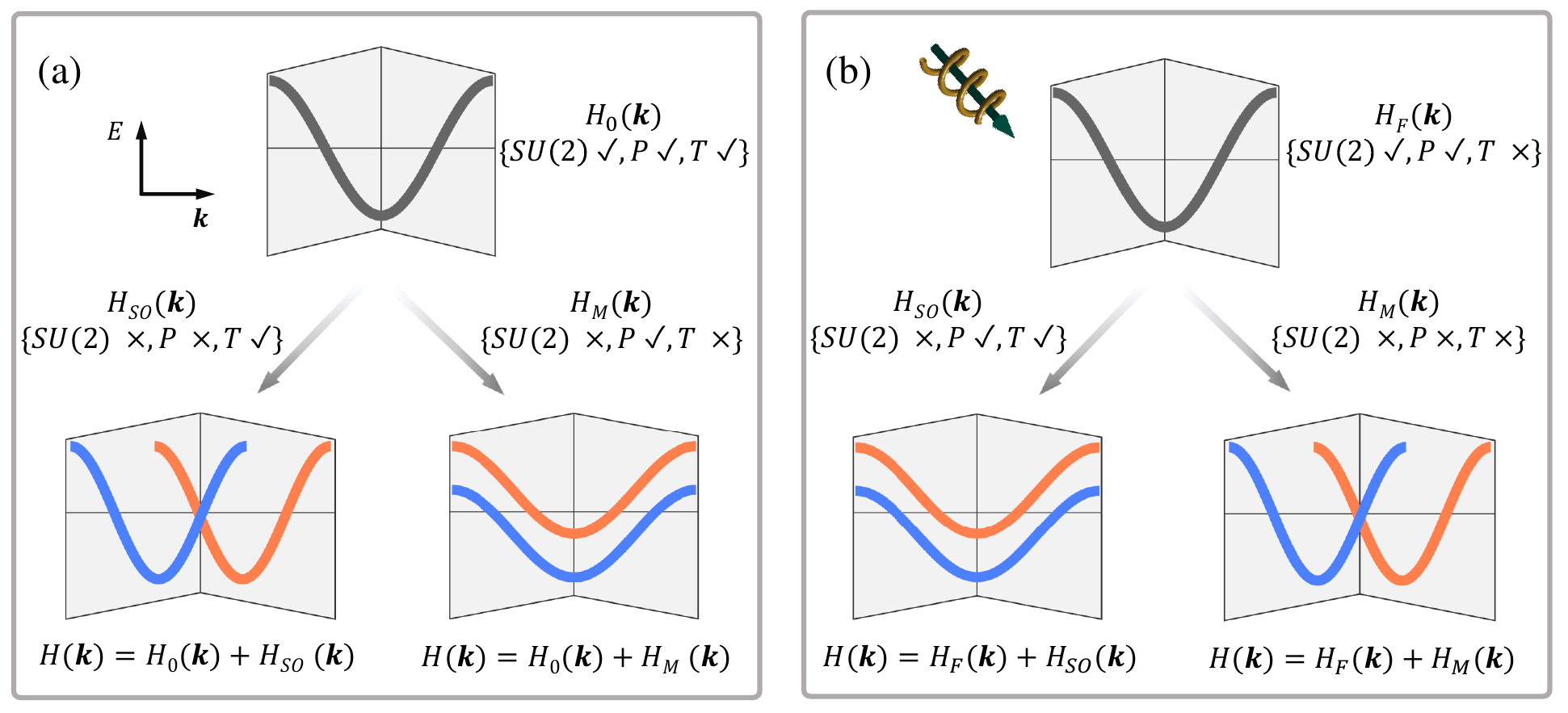}
\caption{(a) Conventional picture of spin splitting induced by SOC and MEC. 
Starting from a static Hamiltonian $H_{0}$ that preserves spin-SU(2), $\mathcal{T}$ and $\mathcal{P}$, the inclusion 
of $\mathcal{T}$-invariant and $\mathcal{P}$-broken SOC ($H_{\rm SO}$) leads to odd-parity spin splitting, 
whereas the inclusion of  $\mathcal{T}$-broken and $\mathcal{P}$-invariant MEC ($H_{\rm M}$) leads to even-parity spin splitting.
(b) A counterpart to the conventional picture. Starting from a CPL-driven Hamiltonian $H_{F}$ that preserves spin-SU(2) and $\mathcal{P}$ but lacks 
$\mathcal{T}$, the inclusion 
of $\mathcal{T}$-invariant and $\mathcal{P}$-invariant SOC  leads to even-parity spin splitting, 
whereas the inclusion of  $\mathcal{T}$-broken and $\mathcal{P}$-broken MEC (preserving the combination of 
a $\pi$ spin rotation and a spatial inversion) leads to odd-parity spin splitting. The symbol ``$\surd$'' indicates 
that the symmetry is preserved, while ``$\times$'' means the symmetry is broken. 
}\label{fig1}
\end{figure}

Spin splitting is a cornerstone of spintronics~\cite{Wolf2001,Sarma2004spintronics} and underpins the realization of numerous topological 
phases of matter~\cite{Fu2008TSC,Sato2009TSC,Lutchyn2010,Oreg2010,Yu2010QAHE,Chang2013QAHE}. In solid-state systems, it is predominantly generated by two microscopic mechanisms: 
relativistic spin-orbit coupling (SOC) and nonrelativistic magnetic exchange coupling (MEC). While both mechanisms break the 
spin-SU(2) symmetry, a more fundamental distinction lies in 
how they each break the composite $\mathcal{PT}$ symmetry (the product of parity $\mathcal{P}$ 
and time-reversal $\mathcal{T}$). This distinct symmetry-breaking pathway leads to spin splitting 
with qualitatively different characteristics.
To elucidate this, note that, if SOC and MEC are both ignored, a typical equilibrium system is commonly 
described by a spinless Hamiltonian $H_{0}$ that preserves spin-SU(2), $\mathcal{P}$, and $\mathcal{T}$.
Because SOC preserves $\mathcal{T}$, introducing it to break $\mathcal{PT}$
necessitates breaking $\mathcal{P}$. The unbroken $\mathcal{T}$ then mandates 
an odd-parity spin splitting~\cite{Galitski2013,Manchon2015} [Fig.~\ref{fig1}(a), left].
MEC, by contrast, intrinsically breaks $\mathcal{T}$. In magnetic systems that preserve $\mathcal{P}$---such as collinear ferromagnets and the recently discovered altermagnets~\cite{Libor2022AMa,Libor2022AMb,Wu2007Fermi,Hayami2019AM,Hayami2020AM,Yuan2020AM,
Yuan2021AM,Mazin2021,Shao2021NC,Ma2021AM,Liu2022AM}---$\mathcal{PT}$ is broken via time-reversal violation. The remaining $\mathcal{P}$-symmetry then enforces an even-parity spin splitting [Fig.~\ref{fig1}(a), right]. This parity-based dichotomy provides a fundamental classification, distinguishing spin splitting arising from noncentrosymmetric systems with SOC from that inherent to collinear magnetic orders.

This conventional picture, however, rests on the assumption that the SU(2)-invariant part of the total Hamiltonian preserves both 
$\mathcal{P}$ and $\mathcal{T}$. Although this holds for many static systems, a broad range of static and dynamic systems fall outside this scenario. Two representative classes are static systems with loop current orders~\cite{Haldane1988,Bourges2021loop,Fernandes2026} and Dirac materials driven by circularly polarized light (CPL)~\cite{Oka2019review,Rudner2020review,Bao2022review,Zhan2024review}. In such systems, the spin-independent part of the Hamiltonian lacks $\mathcal{T}$-symmetry. Notably, recent theoretical work has shown that odd-parity spin splitting can emerge in a collinear antiferromagnet with 
loop current order~\cite{Lin2025OAM,Zeng2025OAM}, complex orbital order~\cite{Zhuang2025OPAM}, or driven by CPL~\cite{Huang2026OPAM,Li2026OPAM,Liu2026OPAM,Pan2025OPAM,Zhu2026OPAM,Tian2026OPAM} [Fig.~\ref{fig1}(b), right], leading to the discovery of a distinct class of altermagnets---termed odd-parity altermagnets. This finding reveals a key insight: 
breaking the $\mathcal{T}$-symmetry of the SU(2)-invariant part of the Hamiltonian enables a reversal of 
the spin-splitting parity in magnetic systems~\cite{Zhu2026Floquet}.

In this work, we demonstrate that even-parity spin-splitting can emerge in nonmagnetic SOC systems. 
To be specific, we consider centrosymmetric systems where SOC preserves both $\mathcal{T}$ and $\mathcal{P}$. 
In the static case, the band structure of the system is spin-degenerate due to $\mathcal{PT}$ symmetry. 
However, when the system is subjected to CPL, 
the $\mathcal{PT}$ symmetry is broken due to the lifting of the $\mathcal{T}$-symmetry, and 
the remaining $\mathcal{P}$-symmetry forces the emergent spin splitting to be even parity [Fig.~\ref{fig1}(b), left]. 
Using minimal two-orbital models, we show that spin-split band structures reminiscent of those in 
$s$-wave ferromagnets, $d$-wave altermagnets, and $g$-wave altermagnets can be achieved. Remarkably, 
the symmetry of the resulting spin splitting is 
governed by the angular character of the underlying orbitals. This work, together with previous 
studies on odd-parity altermagnets, establishes a counterpart to the conventional picture of spin splitting 
induced by SOC and MEC. Meanwhile, it reveals a route to generating phenomena previously thought to be 
unique to magnetic systems in nonmagnetic materials.

{\it Light-induced $s$-wave spin splitting.---}We focus on two-dimensional systems to illustrate how even-parity spin splitting is induced by CPL in centrosymmetric SOC systems, and how the symmetry of the resulting spin splitting is connected to the angular character of the underlying orbitals. We first consider a square lattice system in which each site hosts both $p_{x}$
and $p_{y}$ orbitals. In the basis $\psi_{\bk}^{\dag} = (c_{p_{x},\uparrow,\bk}^{\dag}, c_{p_{x},\downarrow,\bk}^{\dag}, c_{p_{y},\uparrow,\bk}^{\dag}, c_{p_{y},\downarrow,\bk}^{\dag})$, the Hamiltonian reads $H = \sum_{\bk} \psi_{\bk}^{\dag} \mathcal{H}(\bk) \psi_{\bk}$, where $\mathcal{H}(\bk)=\mathcal{H}_{0}(\bk)+\mathcal{H}_{\rm SO}(\bk)$ with
\begin{eqnarray}
\mathcal{H}_{0}(\bk) &=& 2t_{1} (c_{x} + c_{y}) + 2t_{2} (c_{x} - c_{y})\tau_{z} + 4t_{3} s_{x} s_{y}\tau_{x},\nonumber\\
\mathcal{H}_{\rm SO}(\bk)&=&\lambda \tau_{y} \sigma_{z}. \label{swave}
\end{eqnarray}
Here we use the shorthand notation $c_{x/y}\equiv \cos k_{x/y}$ and $s_{x/y}\equiv\sin k_{x/y}$
(lattice constants are set to unity).  The Pauli matrices $\tau_{i}$ and $\sigma_{i}$ act on 
orbital and spin degrees of freedom, respectively (identity matrices $\tau_{0}$ and $\sigma_{0}$ are omitted throughout). 
$\mathcal{H}_{0}$ is the spin-SU(2)-invariant part, while 
$\mathcal{H}_{\rm SO}$ represents an onsite SOC term that is both $\mathcal{T}$-invariant and $\mathcal{P}$-invariant.  
The first two terms in $\mathcal{H}_{0}$ together describe anisotropic intraorbital hoppings, and 
the third term accounts for interorbital hopping.  
The SOC reduces the spin-SU(2) symmetry to a spin-U(1) symmetry. 
This residual U(1) symmetry ensures that the $z$-component of spin remains a good quantum number, 
allowing the Hamiltonian to be decomposed into two independent spin blocks: $\mathcal{H}(\bk) = \mathcal{H}_{\uparrow}(\bk) \oplus \mathcal{H}_{\downarrow}(\bk)$.
Because the full Hamiltonian is invariant under both inversion ($\mathcal{P}=\text{I}$) and time-reversal ($\mathcal{T}=i\sigma_{y}\mathcal{K}$), its energy spectrum is doubly degenerate. 
This implies the absence of spin splitting in centrosymmetric systems despite the presence of SOC. This behavior contrasts with that of noncentrosymmetric systems, where SOC preserves $\mathcal{T}$ but breaks $\mathcal{P}$, leading to spin splitting~\cite{Dresselhaus1955SOC,winkler2003spin}.

The spin-degenerate energy spectrum reads
\begin{eqnarray}
E_{\pm}(\bk) = 2t_{1} (c_{x} + c_{y}) \pm \sqrt{ 4t_{2}^{2} (c_{x} - c_{y})^{2} + 16t_{3}^{2}s_{x}^{2}s_{y}^{2} + \lambda^{2}}.\nonumber\\
\end{eqnarray}
In the absence of SOC ($\lambda = 0$), the spectrum exhibits two 
fourfold band degeneracies at $\boldsymbol{\Gamma} = (0,0)$ and $\textbf{M} = (\pi,\pi)$. 
These degeneracies are Dirac points with a quadratic dispersion, analogous to those found
 at the $\textbf{K}$ and $\textbf{K}'$ valleys of bilayer graphene~\cite{McCann2006BG,Novoselov2006BG}. 
The SOC term acts as a Dirac mass, gapping out both quadratic Dirac 
points and resulting in a fully gapped band structure.

The band structure of a Dirac material can be significantly modulated by coupling with 
CPL. In particular, when the pristine Dirac material preserves $\mathcal{T}$, CPL 
can explicitly break it through the generation of $\mathcal{T}$-broken Dirac mass terms~\cite{Oka2009,Gu2011,Perez2014,cayssol2013floquet}. 
This not only strongly alters the transport behavior of the system via the change in symmetry~\cite{McIver2020,Naotaka2025AHE}, 
but also provides a versatile approach to 
inducing topological phase transitions~\cite{Inoue2010,wang2014floquet,yan2016tunable,Chan2016type,Narayan2016,Hubener2017,Yan2017Weyl,Ezawa2017FWSM,
Li2019Floquet,Wang2023FWSM,Liu2026Weyl,Ghorashi2025SOC}. In what follows, we investigate 
the influence of CPL on the Dirac band structure 
of this system.

We consider CPL incident along the $z$ direction, described by the vector 
potential $\bA(t)=A_{0}(\cos\omega t, \eta\sin\omega t)$. Here, $A_{0}$ is 
the amplitude, $\omega$ is the frequency, and $\eta=\pm1$ denotes the helicity.
Using the Peierls substitution, $\bk\rightarrow \bk+e\bA(t)/\hbar$ (for notational simplicity, 
hereafter we set $e=\hbar=1$), the light-dressed Hamiltonian becomes time-periodic and admits a frequency expansion: $\mathcal{H}(\bk,t)=\sum_{n}\mathcal{H}^{(n)}(\bk)e^{in\omega t}$. We consider 
the high-frequency, off-resonant regime defined by $\omega > W$, with $W$ denoting the bandwidth. 
In this regime, the Floquet-Bloch theory shows that the quasienergy spectrum is effectively captured 
by a static Hamiltonian~\cite{Kitagawa2011Floquet,Goldman2014}
\begin{eqnarray}
\mathcal{H}_{\rm eff}(\bk)=\mathcal{H}^{(0)}+\sum_{n\geq 1}\frac{[\mathcal{H}^{(n)},\mathcal{H}^{(-n)}]}{n\omega}+\mathcal{O}(\omega^{-2}).\label{Floquet}
\end{eqnarray} 
Truncating the summation to $n=2$, we obtain $\mathcal{H}_{\rm eff}(\bk)=\mathcal{H}_{\rm F}(\bk)+\mathcal{H}_{\rm SO}$, where 
$\mathcal{H}_{\rm SO}(\bk)$ retains its form in Eq.~(\ref{swave}) and the SU(2)-invariant part $\mathcal{H}_{\rm F}(\bk)$ reads
\begin{eqnarray}
\mathcal{H}_{\rm F}(\bk)&=&2J_{0}(A_{0})t_{1} (c_{x} + c_{y})+2J_{0}(A_{0})t_{2} (c_{x} - c_{y})\tau_{z}\nonumber\\
&&+4J_{0}(\sqrt{2}A_{0})t_{3} s_{x} s_{y}\tau_{x}+[F_{s1}(\bk)+F_{s2}(\bk)]\tau_{y},\quad
\end{eqnarray}
The function $J_{n}(x)$ denotes the $n$th-order Bessel function of the first kind. 
The most significant consequence of the drive is the appearance of the last $\tau_{y}$-term
which breaks $\mathcal{T}$ while preserving $\mathcal{P}$. This term, together with the SOC term, 
leads to the emergence of even-parity spin splitting.  
The explicit forms of the functions $F_{sn}(\bk)$ with $n=1,2$---where the 
subscript `$s$' indicates $s$-wave symmetry and the subscript `$n$' indicates that the term arises from an $n$-photon process---are
\begin{eqnarray}
F_{s1}(\bk)&=&-\frac{16\sqrt{2}\eta}{\omega}t_{2}t_{3}J_{1}(A_{0})J_{1}(\sqrt{2}A_{0})(s_{x}^{2}c_{y} +s_{y}^{2}c_{x})\nonumber\\
F_{s2}(\bk)&=&-\frac{16\eta}{\omega}t_{2}t_{3}J_{2}(A_{0})J_{2}(\sqrt{2}A_{0})(c_{x}^{2}c_{y} +c_{y}^{2}c_{x}).\quad
\end{eqnarray}
These two functions share two key properties. First, they are invariant 
under $C_{4z}$ rotation. Second, they change sign under a $(\pi,\pi)$ momentum shift and exhibit nodes 
along the lines $k_{x}=\pm k_{y}\pm\pi$ in the Brillouin zone. The symmetry and the nodal structure 
are analogous to that of an extended $s$-wave pairing described by the gap 
function $(\cos k_{x}+\cos k_{y})$ in superconductors~\cite{Hirschfeld2011}. Consequently, upon 
a more refined classification, the symmetry of the light-induced spin splitting can be identified as extended $s$-wave~\cite{Matteo2026swave}.

\begin{figure}[t]
\centering
\includegraphics[width=0.45\textwidth]{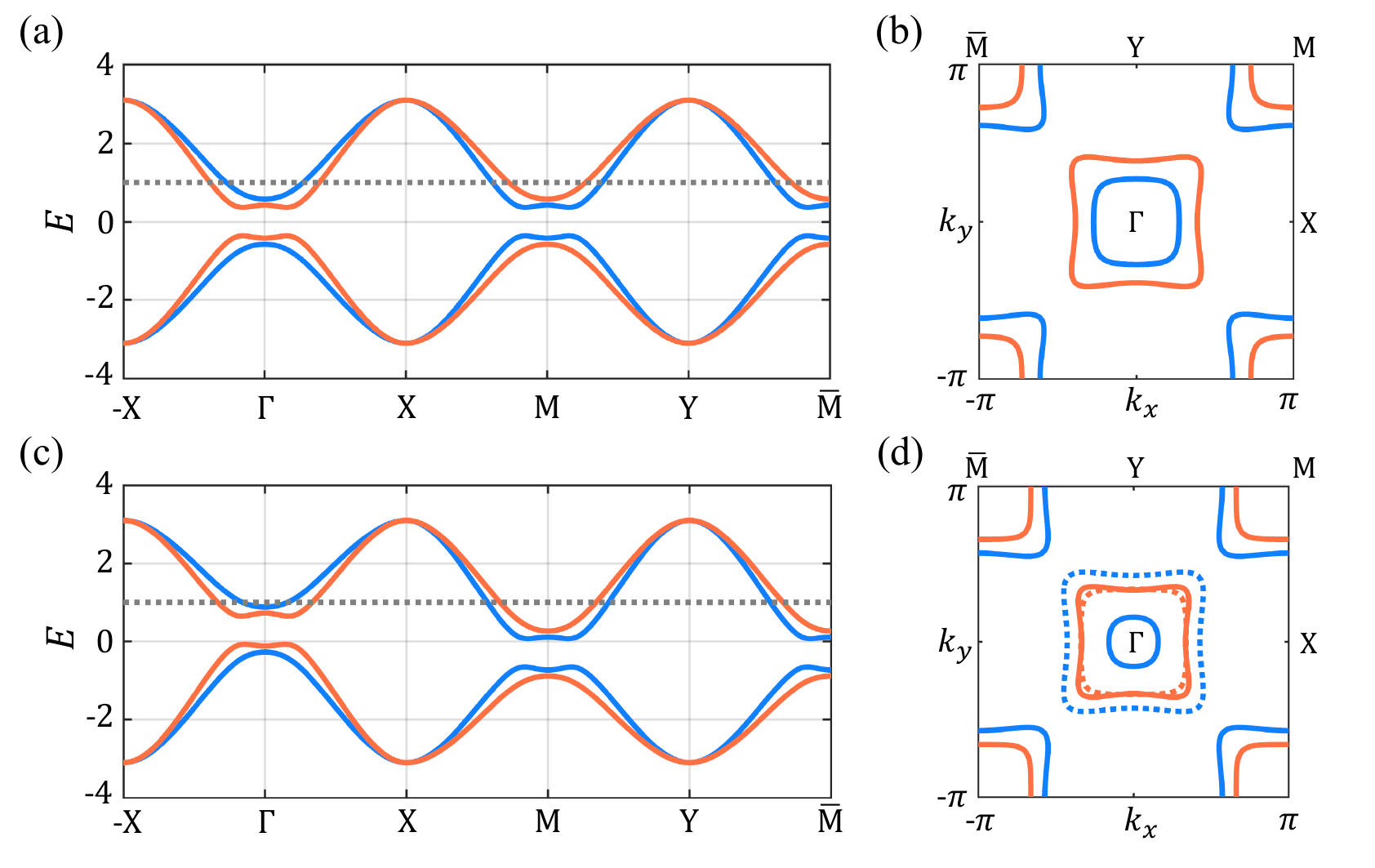}
\caption{Light-induced extended $s$-wave spin splitting. Energy bands along high-symmetry lines
for the conduction-valence-band symmetric [(a), $t_{1}=0$] and asymmetric [(c), $t_{1}=0.1$] case, respectively. 
Orange and blue lines represent spin-up and spin-down bands, respectively. 
Panels (b) and (d) show the isoenergy contours at $E=1$, with (b) [(d)] corresponding to (a) [(c)]. 
In (d), the dashed rings corresponds to a 
$(\pi,\pi)$ shift of the solid rings centering round the $\textbf{M}$ point. Shared parameters 
are: $t_{2}=1$, $t_{3}=0.5$, $\lambda=0.5$, $\omega=5$, $\eta=1$, and $A_{0}=1.0$.
}\label{fig2}
\end{figure}

In Figs.~\ref{fig2}(a) and \ref{fig2}(c), we show the spin-split energy spectrum along high-symmetry 
lines in the Brillouin zone for both the conduction-valence-band symmetric ($t_{1}=0$) 
and asymmetric ($t_{1}\neq0$) cases. The spectra along the lines 
$-\textbf{X}$-$\boldsymbol{\Gamma}$-$\textbf{X}$ and $\textbf{M}$-$\textbf{Y}$-$\bar{\textbf{M}}$ clearly 
reveal that the spin splitting is momentum-dependent and exhibits even parity. For 
the symmetric case, the energy bands satisfy 
$E_{\uparrow}(k_{x},k_{y})=E_{\downarrow}(k_{x}+\pi,k_{y}+\pi)$. Consequently, the isoenergy 
contours for spin-up and spin-down electrons always have equal size [see Fig.~\ref{fig2}(b)], 
leading to zero net spin magnetization regardless of the filling. In the asymmetric case, 
this perfect compensation is lifted [see Fig.~\ref{fig2}(d)], giving rise to a finite spin 
magnetization when the system is metallic.

\begin{figure}[t]
\centering
\includegraphics[width=0.45\textwidth]{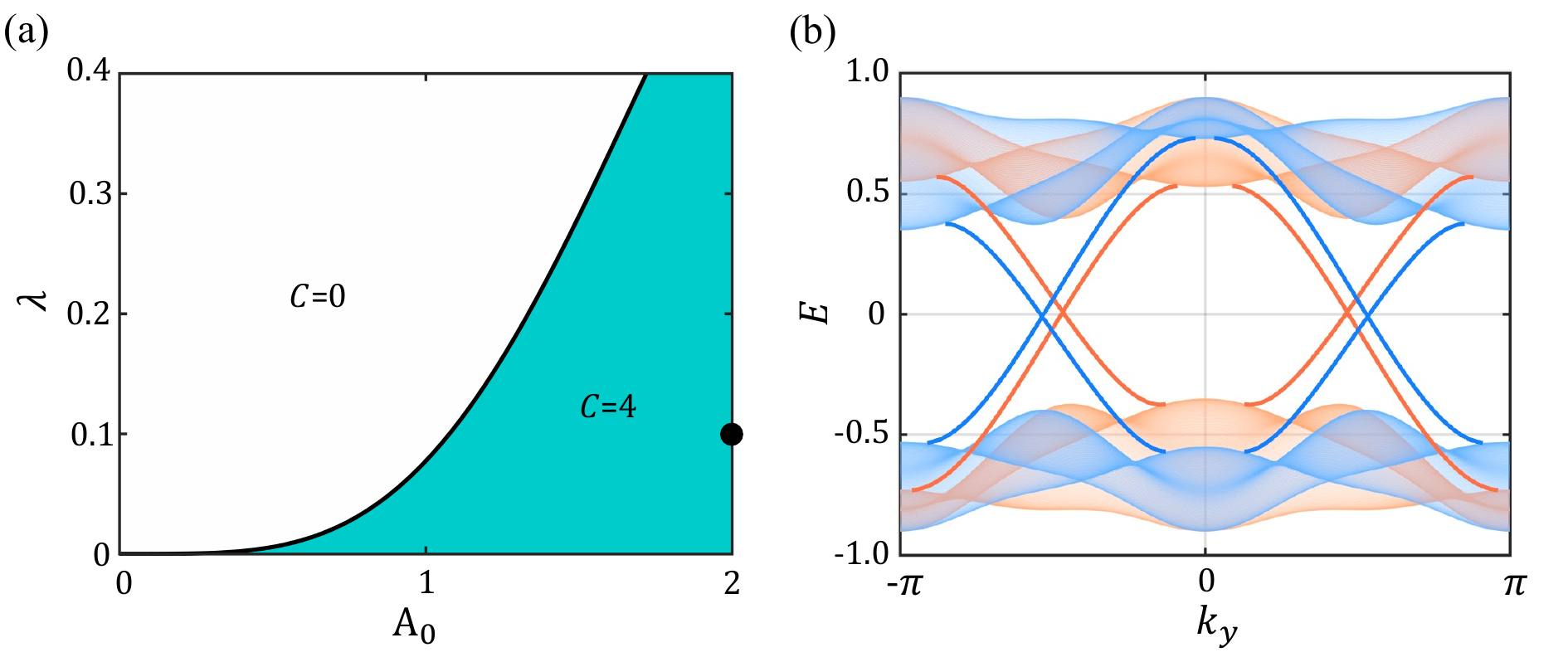}
\caption{(a) Topological phase diagram.
(b) Open-boundary spectrum for the Chern insulator phase labeled by
the black dot illustrated in panel (a). Each edge contains two 
spin-up and two spin-down chiral edge states, consistent with the Chern number $C=4$. 
Shared parameters are: $t_{1}=0.1$, $t_{2}=1$, $t_{3}=0.5$,  $\omega=5$, and $\eta=1$.
}\label{fig3}
\end{figure}

Owing to the underlying Dirac band structure, the spin-split bands become 
topologically nontrivial when the driving amplitude is sufficiently strong 
to close the energy gap at the $\boldsymbol{\Gamma}$ and $\textbf{M}$ points. Since $F_{s1}$ vanishes 
at these points while $F_{s2}$ remains finite, the gap closure is governed 
by $F_{s2}$ and occurs when $\lambda = |F_{s2}(\boldsymbol{\Gamma}/\textbf{M})|$ (note that $|F_{s2}(\boldsymbol{\Gamma})| = |F_{s2}(\textbf{M})|$). 
The resulting topological phase diagram is shown in Fig.~\ref{fig3}(a). Interestingly, 
a simultaneous gap closure at both $\boldsymbol{\Gamma}$ and $\textbf{M}$ enables the 
realization of a Chern insulator phase with Chern number $C = 4$. The bulk-boundary correspondence  
dictates the presence of two spin-up and two spin-down gapless chiral edge states on the boundary, 
as shown in Fig.~\ref{fig3}(b). 

It is instructive to compare the spin-split band structure of this light-induced phase with that of a conventional ferromagnet. 
Although both phases exhibit $s$-wave spin splitting, 
the light-induced phase is characterized by momentum-dependent spin splitting and nodal lines where the 
splitting vanishes, whereas ferromagnetic splitting is momentum-independent and nodeless. Therefore, this 
light-induced phase can be regarded as an unconventional itinerant ferromagnetic phase without localized 
magnetic moments.

{\it Light-induced d-wave spin splitting.---}The zero net magnetization in the extended $s$-wave spin-split phase relies on the condition $E_{\uparrow}(k_{x},k_{y})=E_{\downarrow}(k_{x}+\pi,k_{y}+\pi)$. This condition, however, is no longer required when the spin splitting belongs to a higher partial-wave channel. To demonstrate this, we consider a square lattice model with two orbitals per site: $d_{x^{2}-y^{2}}$ and $d_{z^{2}}$. In the basis $\psi_{\bk}^{\dag}=(c_{x^{2}-y^{2},\uparrow,\bk}^{\dag},c_{x^{2}-y^{2},\downarrow,\bk}^{\dag}, c_{z^{2},\uparrow,\bk}^{\dag},c_{z^{2},\downarrow,\bk}^{\dag})$, the Hamiltonian reads $H=\sum_{\bk}\psi_{\bk}^{\dag}\mathcal{H}(\bk)\psi_{\bk}$, 
where 
\begin{eqnarray}
\mathcal{H}(\bk)&=&2t_{+}(c_{x}+c_{y})+2t_{-}(c_{x}+c_{y})\tau_{z}\nonumber\\
&&+2t_{3}(c_{x}-c_{y})\tau_{x}+\lambda \tau_{y}\sigma_{z}.\label{dwave}
\end{eqnarray}
Here, $t_{+}=(t_{x^{2}-y^{2}}+t_{z^{2}})/2$ and $t_{-}=(t_{x^{2}-y^{2}}-t_{z^{2}})/2$, with $t_{x^{2}-y^{2}}$ and $t_{z^{2}}$ denoting the intraorbital hopping amplitudes, and $t_{3}$ the interorbital hopping amplitude. Similar to the $(p_{x},p_{y})$ case, the Hamiltonian in Eq. (\ref{dwave}) preserves both $\mathcal{T}$ and $\mathcal{P}$, resulting in a spin-degenerate band structure. In contrast to the previous case, where quadratic Dirac cones appear at two time-reversal-invariant momenta, this Hamiltonian hosts four linear Dirac cones at $(k_{x},k_{y})=(\pm\pi/2,\pm\pi/2)$ in the absence of SOC ($\lambda=0$).

Applying the same CPL driving and following the Floquet formalism in Eq.~(\ref{Floquet}), we obtain the effective Hamiltonian for the driven system:
\begin{eqnarray}
\mathcal{H}_{\rm eff}(\bk)&=&2J_{0}(A_{0})t_{+}(c_{x}+c_{y})+2J_{0}(A_{0})t_{-}(c_{x}+c_{y})\tau_{z}\nonumber\\
&&+2J_{0}(A_{0})t_{3}(c_{x}-c_{y})\tau_{x}+F_{d1}(\bk)\tau_{y}+\lambda \tau_{y}\sigma_{z},\quad
\end{eqnarray}
where  $F_{d1}(\bk)=-32J_{1}^{2}(A_{0})\eta t_{-} t_{3}s_{x}s_{y}/\omega$ (the two-photon process $[\mathcal{H}^{(2)},H^{(-2)}]$ does not contribute 
in this model).
The driving-induced term $F_{d1}(\bk)\tau_{y}$ also breaks $\mathcal{T}$ while preserving $\mathcal{P}$. 
A key feature of this term  is its sign change under a $\pi/2$ rotation, a character of $d$-wave symmetry.

\begin{figure}[t]
\centering
\includegraphics[width=0.45\textwidth]{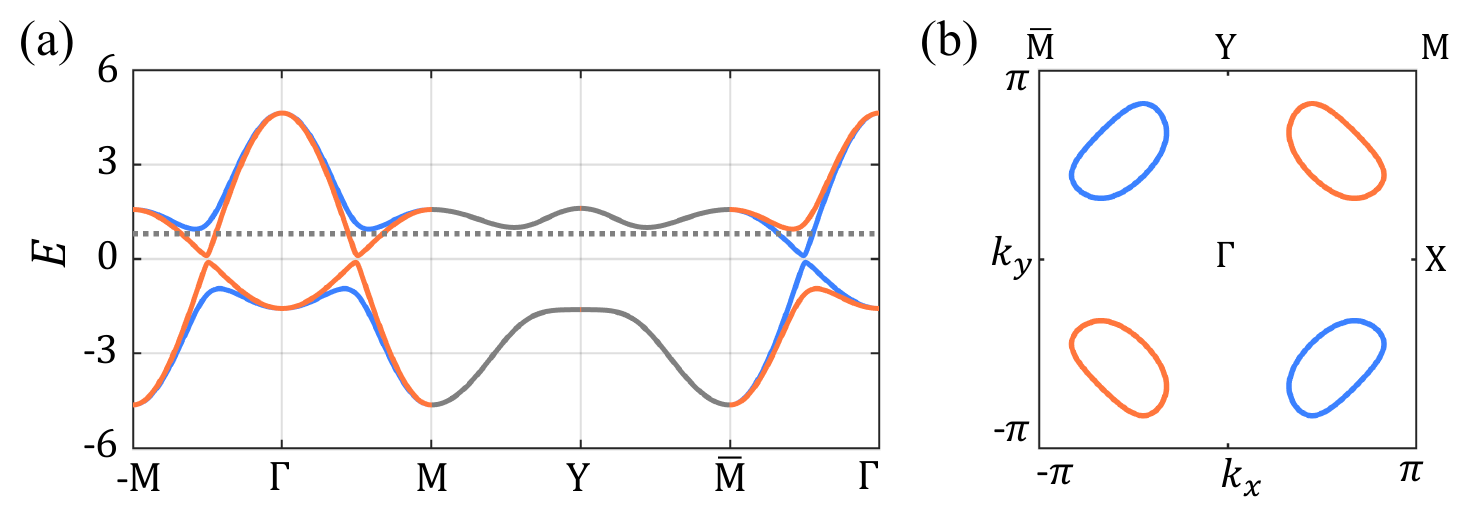}
\caption{ Light-induced $d$-wave spin splitting. (a) Energy bands along high-symmetry lines. Orange and blue curves denote spin-up and spin-down bands, respectively; grey curves indicate spin-degenerate bands. (b) Isoenergy contours at $E=0.8$. The alternating orange and blue lobes reveal a $d_{xy}$-wave symmetry. Such a spin texture is a hallmark of $d$-wave altermagnetism, here induced without intrinsic magnetism. Parameters are:
$t_{+}=0.5$, $t_{-}=1$, $t_{3}=0.5$, $\lambda=0.5$, $\omega=5$, $\eta=1$, and $A_{0}=1$. 
}\label{fig4}
\end{figure}

In Fig.~\ref{fig4}(a), we show the energy spectrum along selected high-symmetry lines. The spin splitting clearly exhibits even parity. 
Figure \ref{fig4}(b) displays the isoenergy contours at a generic energy. It is evident that the spin texture forms a 
$d_{xy}$-wave pattern, analogous to those found in $d$-wave altermagnets~\cite{Libor2022AMb}. This $d$-wave symmetry of 
the spin splitting directly implies zero net spin magnetization.

The light-induced spin-split bands also become topologically nontrivial once the driving amplitude exceeds a critical threshold determined by the SOC strength. Since the energy gap can close only at  $(k_{x},k_{y})=(\pm\pi/2,\pm\pi/2)$, the phase boundary is simply determined by $\lambda = 32t_{-}t_{3}J_{1}^{2}(A_{0})/\omega$. As in the previous case, the topological phase is a Chern insulator, characterized by a Chern number $C=4$ and hosting two spin-up and two spin-down chiral edge states on the boundary.

Before proceeding, we emphasize that spin splittings with even higher partial-wave symmetry can 
be achieved similarly by an appropriate choice of the orbital basis. For instance, a $g$-wave 
spin splitting can be obtained if the basis consists of $d_{xy}$ and $d_{x^{2}-y^{2}}$ orbitals. 
In this basis, the first two terms and the SOC term in Eq.~(\ref{dwave}) 
retain the same form, while the third term changes from the $d$-wave symmetry form 
$(c_{x} - c_{y})$ to the $g$-wave symmetry form $(c_{x} - c_{y})s_{x}s_{y}$. 
Applying CPL will consequently induce a $g$-wave spin splitting (see the Supplemental Material).

{\it Light-induced spin and orbital magnetizations.---}In a collinear magnetic system without SOC, MEC can induce spin 
magnetization but cannot generate orbital magnetization. Here, the coupling of CPL to Dirac band structure results in 
orbital magnetization since it makes two orbital motions related by $\mathcal{T}$ no longer symmetric. This implies 
that the simultaneous emergence of spin and orbital magnetizations would be a generic phenomenon in these 
driven SOC systems if there is no symmetry that forces zero net spin magnetization.

\begin{figure}[t]
\centering
\includegraphics[width=0.45\textwidth]{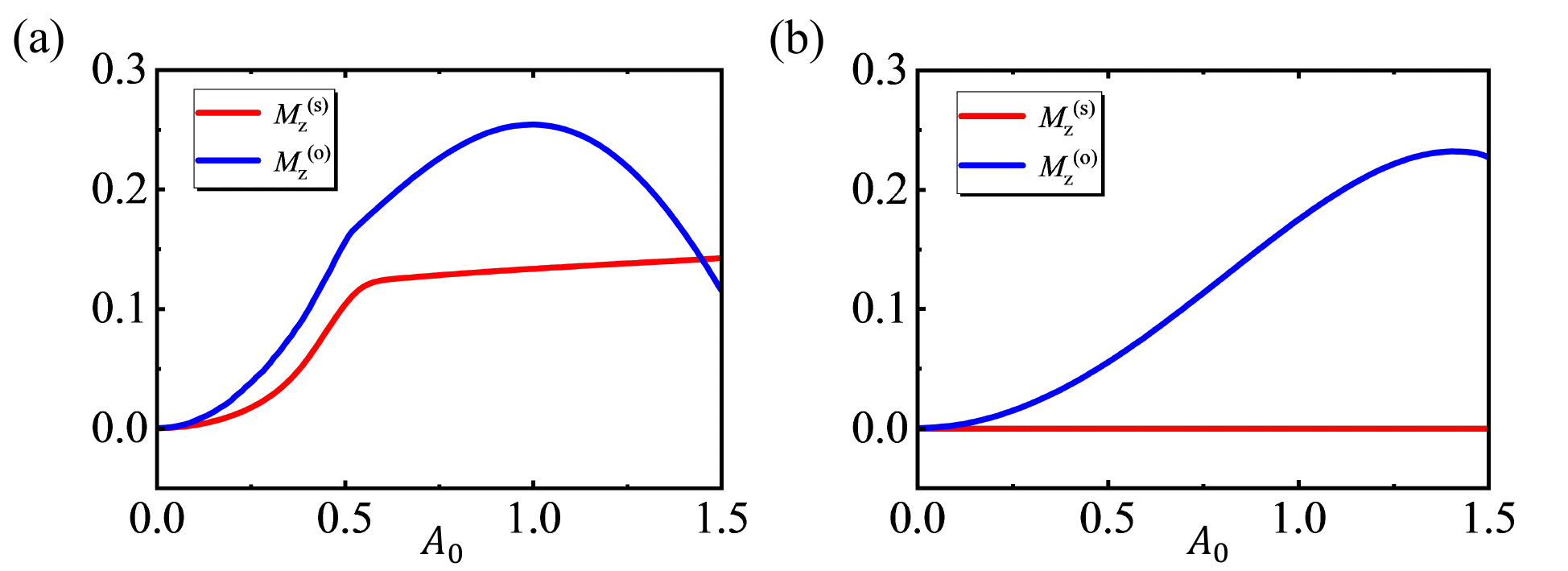}
\caption{ Evolution of spin  and orbital magnetizations  as functions of the driving amplitude. The electron filling fraction 
is fixed at 1. (a) Evolution in the $s$-wave spin-split phase. (b) Evolution in the $d$-wave spin-split phase. The unit of $M_{z}^{(s)}$ is 
$\mu_{B}$, and the unit of $M_{z}^{(o)}$ is $\mu_{B}^{*}=\hbar e/2m^{*}$, where $m^{*}$ is the effective mass defined at the band edge 
of the lower band before driving. Parameters for (a): $\{t_{1},t_{2},t_{3},\lambda,\omega,\eta,A_{0}\}=\{0.5,1,0.5,0.5,5,1,1\}$. 
Parameters for (b): $\{t_{+},t_{-},t_{3},\lambda,\omega,\eta,A_{0}\}=\{0.5,1,0.5,0.5,5,1,1\}$. 
}\label{fig5}
\end{figure}

To determine the spin and orbital magnetizations of the driven system, one would in principle need to obtain 
the distribution function of the nonequilibrium steady state, which depends on the microscopic details of dissipation~\cite{Dehghani2014,Dehghani2015,Esin2018steady,Seetharam2015}. Here, we restrict ourselves to a qualitative understanding by assuming that the off-resonantly driven system remains near equilibrium, 
such that the Fermi-Dirac distribution remains a good approximation. This assumption is further motivated by the fact that the 
same light-driven spin-split phases can also emerge as static phases when the $\mathcal{T}$-symmetry is broken 
by an interaction-driven loop current order~\cite{Sun2009QBC}. 
In such a static realization, the distribution is exactly the Fermi-Dirac distribution.
 
The spin magnetization is calculated via the formula 
\begin{eqnarray}
M_{z}^{(s)} = \mu_{\rm B} \sum_{n} \int_{\bk} \left[ f(E_{n\uparrow}) - f(E_{n\downarrow}) \right],
\end{eqnarray}
where $\int_{\bk}\equiv\int \frac{d^{2}k}{(2\pi)^{2}}$,  $n$ labels the band, $\mu_{\rm B}$ is the Bohr magneton, and $f(E_{n\alpha}) = \left[ e^{(E_{n\alpha} - \mu)/k_{B}T} + 1 \right]^{-1}$ is the Fermi–Dirac distribution function. Here, $\mu$ denotes the chemical potential, $k_{B}$ the Boltzmann constant, and $T$ the temperature.

The orbital magnetization at zero temperature is determined by~\cite{Xiao2005Berry,Xiao2007Berry} 
\begin{eqnarray}
M_{z}^{(o)}=\sum_{n,\alpha}\int_{\bk}[m_{n\alpha}(\bk)+\frac{e}{\hbar}(\mu-E_{n\alpha}(\bk))\Omega_{n\alpha}(\bk)],
\end{eqnarray}
where the integration is carried out over states below the chemical potential $\mu$. 
$m_{n\alpha}(\bk)=-i\epsilon_{ij}\langle \partial_{k_{i}} u_{n}(\bk)|(H(\bk)-E_{n}(\bk))|\partial_{k_{j}} u_{n}(\bk)\rangle$ is the orbital magnetic moment,  and $\Omega_{n\alpha}(\bk)=i \epsilon_{ij}\langle \partial_{k_{i}} u_{n\alpha} | \partial_{k_{j}} u_{n\alpha} \rangle$ is the Berry curvature. 
In these expressions, $\epsilon_{ij}$ is the antisymmetric tensor, and $|u_{n\alpha}(\bk)\rangle$ is the eigenstate of the $n$th spin-$\alpha$ band at momentum $\bk$. 

Figures \ref{fig5}(a) and \ref{fig5}(b) illustrate the dependence of the zero-temperature spin and orbital magnetizations on the driving amplitude for the 
$s$-wave and $d$-wave spin-split phases, respectively. In both cases, the total electron filling fraction $n_{e} = n_{\uparrow} + n_{\downarrow}$ is fixed at $1$. At $A_{0}=0$, both spin and orbital magnetizations vanish as required by $\mathcal{T}$-symmetry.
As  $A_{0}$ increases, the spin magnetization grows in the 
$s$-wave spin-split phase but remains zero in the $d$-wave spin-split phase. 
In the $s$-wave case, the spin magnetization can reach an order of magnitude of 
0.1$\mu_{B}$. Meanwhile, the orbital magnetization grows monotonically with
$A_{0}$ in the weakly driven regime for both $s$-  and $d$-wave phases. In both phases, 
the orbital magnetization can reach an order of magnitude of $0.2\mu_{B}^{*}$, where the effective Bohr magneton 
$\mu_{B}^{*}=e\hbar/2m^{*}$ is defined in terms of the effective mass at the band edge.

{\it Discussions and conclusion.---}The magnitude of the spin splitting is 
jointly determined by the strength of SOC and the light amplitude. Specifically, 
at a given momentum $\bk$, the spin splitting is bounded by $\Delta E = 2 \min\{\lambda, |F_{D}(\bk)|\}$, 
where $F_{D}(\bk)$ denotes the driving-induced term. Consequently, the maximum achievable spin splitting 
is limited by the SOC strength. This suggests that materials containing heavy elements are preferred 
for realizing a large light-induced spin splitting. Bismuth, in particular, possesses strong atomic SOC and 
its outer-shell electrons occupy $p$ orbitals, making Bi-based materials ideal candidates for exploring the 
light-induced $s$-wave spin-split phase. As a concrete example, we suggest bilayer Bi(111), a nonmagnetic, centrosymmetric 
 topological insulator with strong SOC~\cite{Liu2011Bi,Drozdov2014}. To observe $d$- and $g$-wave spin splittings, we suggest 
transition-metal oxides, which are a large family of materials involving $d$-orbital electrons. 
For example, symmetric bilayer transition-metal oxides with fractionally-filled $e_{g}$ orbitals ($d_{z}^{2}$ and $d_{x^{2}-y^{2}}$), 
e.g., LaAuO$_{3}$, naturally exhibit Dirac band structures when grown along 
the [111] direction~\cite{Xiao2011TMO}, and thus represent concrete material candidates for the $d$-wave spin-split phase. Intriguingly, 
because spontaneous $\mathcal{T}$-symmetry breaking can occur in these materials when interactions drive the formation of complex orbital orders~\cite{Fiete2011}, the predicted spin-split phases may be observable even without optical driving.

In experiments, the most direct way to observe light-induced change in band structure is time-resolved ARPES~\cite{wang2013observation,Zhou2023Floquet,Zhou2013low,Merboldt2025,choi2025,Wang2026Floquet}. 
We emphasize that although our investigation focuses on the high-frequency regime to establish 
an analytical understanding, all essential physics remain the same in the lower-frequency regime, since the 
even-wave spin splitting is enforced by symmetry. In fact, lower frequencies can induce stronger 
spin splitting, as evidenced by the $1/\omega$ dependence of the driving terms. Furthermore, the dynamical generation 
of even-wave spin polarization can be generalized to systems without spin conservation. Even when spin is not conserved---for example, in three-dimensional systems with SOC of the form $\boldsymbol{l}\cdot\boldsymbol{s}$---the remaining inversion 
symmetry is sufficient to force all spin polarization components $\langle s_i \rangle$ to have even parity: $\langle s_i(\bk)\rangle = \langle s_i(-\bk)\rangle$. The specific symmetry of the spin splitting can be detected not 
only by ARPES, through measurements along high-symmetry lines~\cite{Osumi2024MnTe,Lee2024MnTe,Krempasky2024,
Hajlaoui2024AM,Reimers2024,Ding2024CrSb,Yang2024CrSb,
Zeng2024CrSb,Li2024CrSb,Jiang2024KV2Se2O}, but also by STM, through 
quasiparticle interference, as exemplified by recent STM experiments on $d$-wave altermagnetic metals~\cite{Mu2026dwave}.

In summary, a key finding of this work is the demonstration that ferromagnet-like and altermagnet-like 
even-wave spin-split band structures can emerge in nonmagnetic, centrosymmetric systems with SOC, provided that the 
$\mathcal{T}$-symmetry is lifted by CPL. These optically induced phases offer a new pathway to generate 
even-parity spin-split bands without intrinsic magnetism.

{\it Acknowledgements.---}We thank Prof. Shuyun Zhou and Prof. Zhigang Wu for helpful discussions and comments 
on the manuscript. This work is supported by 
Guangdong Basic and Applied Basic Research Foundation (Grant No. 2023B1515040023), 
and Fundamental and Interdisciplinary Disciplines 
Breakthrough Plan of the Ministry of Education of China (JYB2025XDXM403).

\bibliography{dirac.bib}

\appendix 

\section{End Matter}

\renewcommand{\thefigure}{E\arabic{figure}}
\renewcommand{\theequation}{E\arabic{equation}}
\setcounter{equation}{0}
\setcounter{figure}{0}

{\it Light-induced $g$-wave spin splitting.---}We consider a square lattice system where each 
site consists of $d_{x^2-y^2}$ and $d_{xy}$ orbitals. In the basis $\psi_{\bk}^{\dagger}=(c_{x^2-y^2, \uparrow, \bk}^{\dagger}, c_{x^2-y^2 , \downarrow, \bk}^{\dagger},c_{xy, \uparrow, \bk}^{\dagger}, c_{xy, \downarrow, \bk}^{\dagger})$, the Hamiltonian is given by \(H=\sum _{\bk}\psi^\dagger_{\bk}  \mathcal{H}(\bk)\psi_{\bk}\), where 
\begin{eqnarray}
\mathcal{H}(\bk)&=&2t_{+}(c_x+c_y)
+2t_{-}(c_x+c_y)\tau_z\nonumber\\
&&+4t_3s_xs_y(c_x-c_y)\tau_x
+\lambda\tau_y\sigma_z. \label{E1}
\end{eqnarray}
For clarity, we recall the shorthand notation: 
$c_{x/y}=\cos k_{x/y}$ and $s_{x/y}=\sin k_{x/y}$. The two parameters $t_+$
and $t_{-}$ are defined in terms of  intraorbital hopping amplitudes $t_{x^2-y^2}$ and $t_{xy}$:
$t_{+}=(t_{x^2-y^2}+t_{xy})/2$ and $t_{-}=(t_{x^2-y^2}-t_{xy})/2$. The parameter $t_3$ is the interorbital hopping amplitude, 
and $\lambda$ characterizes the strength of spin-orbit coupling (SOC). 

Rather than deriving the full Floquet lattice Hamiltonian as in the main text, we adopt 
a low-energy continuum approach here to better illustrate the emergence of 
$g$-wave spin splitting under circularly polarized light (CPL). Specifically, we expand the Hamiltonian around the high-symmetry momentum 
$\boldsymbol{\Gamma}=(0,0)$ and use this low-energy description to derive the Floquet Hamiltonian. Since the term 
$t_{+}(c_{x}+c_{y})$ does not contribute to the spin splitting, we neglect it for simplicity. The continuum low-energy Hamiltonian around 
$\boldsymbol{\Gamma}=(0,0)$ before driving is given by
\begin{eqnarray}
 		\mathcal{H}(\bq)=t_{-}(4-q^{2})\tau_z
 		-2t_3q_xq_y(q_x^2-q_y^2)\tau_x+\lambda\tau_y\sigma_z,\quad
\end{eqnarray}
where $\bq=(q_{x},q_{y})$ represents the momentum measured from $\boldsymbol{\Gamma}$.

Under illumination by CPL described by the vector potential $\bA(t)=A_0(\cos \omega t,\eta \sin \omega t)$, 
the effect of CPL is captured by the substitution: $\mathcal{H}(\bq)\to \mathcal{H}(\bq+e\bA(t)/\hbar)$ (for notational 
simplicity, below we set $e=\hbar=1$ as in the main text). 
The Hamiltonian becomes time-periodic and can be expanded in Fourier components as $\mathcal{H}(\bq+\bA(t))=\sum_n \mathcal{H}^{(n)}(\bq) e^{in\omega t}$, with
\begin{eqnarray}
		\mathcal{H}^{(0)}(\bq)&=&\, t_{-}(4-q^{2}-A_0^2)\tau_z
		-2t_3q_xq_y(q_x^2-q_y^2)\tau_x\nonumber\\
&&+\lambda\tau_y \sigma_z,\nonumber\\
		\mathcal{H}^{(\pm1)}(\bq)&=& -t_{-}A_0(q_x\mp i\eta q_y)\tau_z\nonumber\\
		&&-t_3A_0\big[3(q_x^2q_y\pm i\eta q_xq_y^2)-(q_y^3\pm i\eta q_x^3)\big]\tau_x,\nonumber\\
		\mathcal{H}^{(\pm2)}(\mathbf q)&=& 
		-t_3A_0^2\big[3q_xq_y\pm \tfrac{3}{2}i\eta (q_y^2-q_x^2)\big]\tau_x,
		\label{H12}
\end{eqnarray}
and $\mathcal H^{(n)}=0$ for $|n|>2$.

\begin{figure}[b]
\centering
\includegraphics[width=0.4\textwidth]{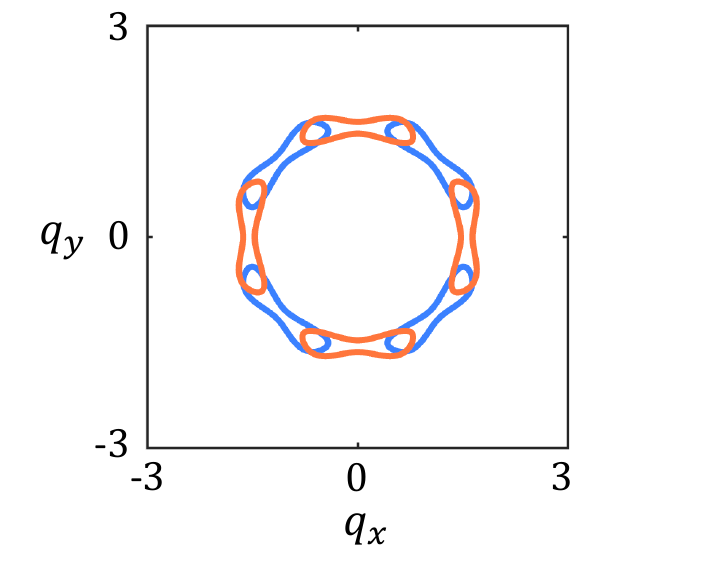}
		\caption{ Light-induced $g$-wave spin splitting. Orange and blue curves denote spin-up and spin-down isoenergy contours at $E=1$, respectively. Parameters are:  $t_{+}=0$, $t_{-}=1$, $t_3=0.2$, $\lambda=0.2$,  $\omega=5$, $\eta=1$, and $A_{0}=1$.}
		\label{g-wave}
	\end{figure}

In the high-frequency off-resonant regime, the quasienergy spectrum is governed by an effective static Hamiltonian
\begin{eqnarray}
	\mathcal{H}_{\rm eff}(\bq)=\mathcal{H}^{(0)}(\bq)+\sum_{n \geq 1} \frac{\left[\mathcal{H}^{(n)}(\bq), \mathcal{H}^{(-n)}(\bq)\right]}{n \omega}+\mathcal{O}\left(\omega^{-2}\right).\nonumber\\
\end{eqnarray}
Using Eq.~(\ref{H12}), we obtain
\begin{equation}
	\begin{aligned}
		[\mathcal H^{(1)},\mathcal H^{(-1)}]
		&=-4\eta t_{-} t_3 A_0^2 \left(q_x^4+q_y^4-6q_x^2 q_y^2\right)\sigma_y,\\
		[\mathcal H^{(2)},\mathcal H^{(-2)}]
		&=0.
	\end{aligned}
\end{equation}
The vanishing commutator $[\mathcal H^{(2)},\mathcal H^{(-2)}]=0$ indicates that the two-photon processes do not contribute to the band renormalization at this order. Consequently, the effective static Hamiltonian reads
\begin{eqnarray}
		\mathcal{H}_{\rm eff}(\bq)
		&=& t_{-}(4-q^2-A_0^2)\tau_z
		-2t_3 q_x q_y (q_x^2-q_y^2)\tau_x \nonumber\\
		&&-\frac{4}{\omega}\eta t_{-} t_3 A_0^2
		(q_x^4+q_y^4-6q_x^2 q_y^2)\tau_y+\lambda\tau_y \sigma_z.\quad\label{Heff}
\end{eqnarray}
For the third term in $\mathcal H_{\rm eff}$ which is induced by the drive, note that
\begin{eqnarray}
q_x^4+q_y^4-6q_x^2 q_y^2=\text{Re}(q_{x}+iq_{y})^{4}=q^{4}\cos 4\theta, 
\end{eqnarray}
which clearly corresponds to the channel with angular momentum $l=4$, i.e., the $g$-wave channel. 
This term, together with SOC, induces a $g$-wave spin splitting. 
In Fig.~(\ref{g-wave}), we present the isoenergy contours of the spectrum of $\mathcal H_{\rm eff}$
at a generic energy; the spin texture explicitly exhibits 
$g$-wave symmetry, analogous to that in $g$-wave altermagnets.

\end{document}